# On The Model Size Selection For Speaker Identification[1]

*Marcos Faúndez-Zanuy*

Escola Universitària Politècnica de Mataró
Universitat Politècnica de Catalunya (UPC)
Avda. Puig i Cadafalch 101-111, E-08303 Mataró (BARCELONA) SPAIN
`faundez@eupmt.es`

## Abstract

In this paper we evaluate the relevance of the model size for speaker identification. We show that it is possible to improve the identification rates if a different model size is used for each speaker. We also present some criteria for selecting the model size, and a new algorithm that outperforms the classical system with a fixed model size.

## 1. Introduction

It is well known that model size selection is a critical fact on pattern recognition, polynomial fitting, etc. If the number of parameters is small, there is not enough precision to model the dates. On the other hand, if the model has a lot of parameters there is an overfit, so the model is unable to generalize and manage mismatch situations.

Although in other fields the model size selection is a well studied problem, little work has been done for speaker identification. Usually, the same model size is used for all the speakers, and this is the unique optimized parameter.

A related subject to the model size selection is the fact that on a biometric systems there are special users considered as "difficult to identify individuals" that require a special treatment. The use of a different model size (or different kind of model, combination of models, etc.) for each speaker is a way to manage these difficult cases.

A second advantage when using different model size for each speaker is that the computational burden during the test phase can be reduced using the minimum model size that lets to recognize a particular speaker.

Using the same database and conditions of our previous work at Eurospeech 99 [1] we will give an example about the relevance on the model size selection and propose to new algorithms inside the next sections.

### 1.1. A simple example

The real optimization situation implies the evaluation of all the possible combinations about model sizes, that is equal to $(N_q)^{Nspeakers}$, where $Nq$ is the number of different model sizes and *Nspeakers* is the number of the speakers in the database. For instance, for eight possible model sizes and *Nspeakers*=49speakers, the result is 1.8e44 different combinations!.

We propose a simple experiment that we think that it is illustrative about model size selection because it lets to plot the results on a bidimensional figure.

This example is not a real situation because it evaluates the identification rates for a given speaker assuming that the remaining speakers have the same model size. This process is repeated as many times as the number of speakers (*Nspeakers*) without considering the model size that has been set for the previous speakers.

*1.1.1. Example of the model size without mismatch.*

The first situation is model size selection without mismatch between training and testing conditions.
This experiment consists on the evaluation of the minimum model size for each speaker that yields the maximum identification rate for this speaker assuming that the remaining speakers have the same model size.

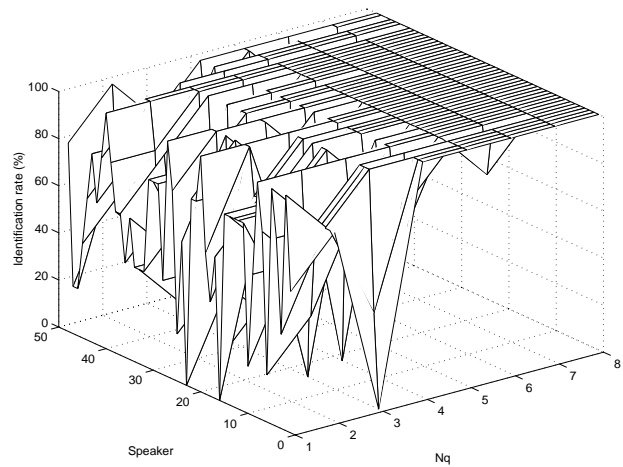

*Figure 1:* Identification rates for each speaker as function of the VQ model size Nq.

Figure 1 shows the identification rates for each speaker as function of the number of bits of a classical VQ identification algorithm [2] named Nq (Number of quantization bits) variable between 0 and 7 bits. Figure 2 shows the histogram with the optimal model sizes.

[1] This work has been supported by the CICYT TIC2000-1669-C04-02

It is easy to check that the grater the model size, the better the identification rates, so there is no improvement in recognition rates choosing a model size smaller that the greatest considered model size. There is only an improvement on the computational burden.

Section 1.1.2 describes a similar situation when there is a mismatch between training and testing conditions.

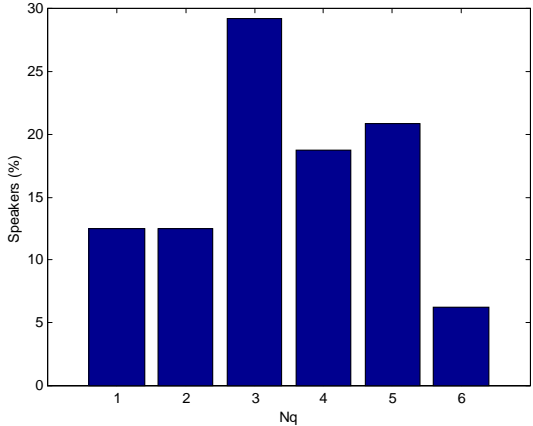

*Figure 2:* histogram of the optimal model size Nq.

*1.1.2. Evaluation of the model size with mismatch.*

In this section we consider a mismatch between training and testing languages. That is, the model of each speaker has been computed in one language, and the test in a different one, with the same conditions than we used on [1].

Figure 3 and 4 show the equivalent results to figure 1 and 2 respectively. In this case, the optimal model size is not always the biggest one. Thus, there is a problem of overfiting, and for several speakers, the greatest size generalizes worse.

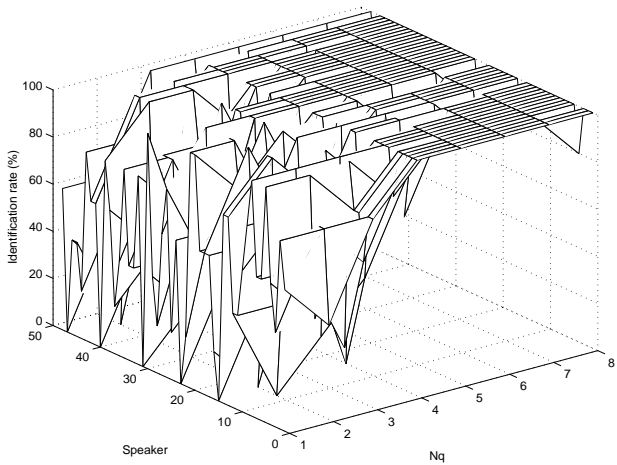

*Figure 3:* Identification rates for each speaker as function of the VQ model size Nq.

According to [3] different training and testing languages is a soft mismatch (the identification rate drops from 100% achieved on section 1.1, to 97.96%), and it is less important than the change of microphone or the use of different recording sessions for enrollment and test. Thus, these results would be more dramatic with a hard mismatch.

The goal of this paper is to find the optimal model size for each speaker, rather than the same optimal fix size for all the speakers, that has been used so far on speaker identification tasks. Next sections describe the proposed algorithms and the achieved results. Section 4 is devoted to the main conclusions of this paper.

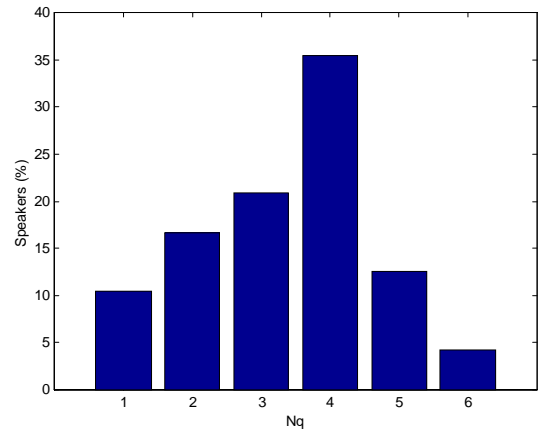

*Figure 4:* histogram of the optimal model size Nq.

## 2. Model size selection using an heuristic procedure

In this section we propose a method to find the optimal model size for each speaker. The full search method requires the evaluation of a tremendously high number of combinations. Even for the situation of only two model sizes, the number of combinations is $2^{49}$=5.6e14. Thus, we propose a practical (and suboptimal) procedure.

### 2.1. Database

Our experiments have been computed over 49 speakers from the Gaudi database [8] that has been obtained with a microphone connected to a PC. The speech signal has been down-sampled to 8kHz, pre-emphasized by a first order filter whose transfer function is $H(z)=1-0.95z^{-1}$. A 30 ms Hamming window is used, and the overlapping between adjacent frames is 2/3. A cepstral vector of order 16 was computed from the LPC coefficients. One minute of read text in Catalan language is used for training, and 5 sentences in Spanish language for testing (each sentence is about 2-3 seconds long).

## 2.2. Proposed algorithm number one.

We propose a suboptimal procedure, that consists on the following steps:
- Initialize all the model sizes $N_{qi}$ $i=1,\cdots N$ equal to a given number of bits, where $N$ is the number of speakers in the database. Thus, all the speakers have a codebook with the same number of bits.

Repeat the following procedure until there is no more improvement over the past iteration:
- The identification rate is evaluated using a test database for $i=1,\cdots N$, with the following conditions: the number of bits of only one speaker is increased to $N_{qi} = N_{qi}+1$, and the others remain the same ($N_{qj} = N_{qj}$, $j=1,\cdots N, j \neq i$).
- It is chosen the $i$ that produces the greatest improvement of the identification rates. The $N_{qi}$ value is incremented in one bit, and this new value is used for this speaker in the next iteration.

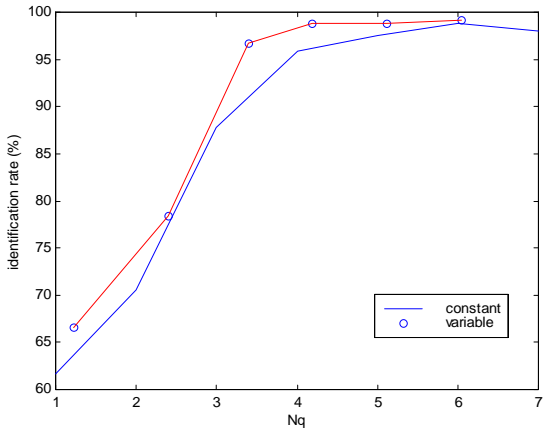

*Figure 5. Identification rates for constant and variable model sizes*

It is interesting to observe that this algorithm presents similarities with the forward selection procedure used to find the optimal components of a vector for pattern recognition applications [5]. The forward selection procedure is an algorithm to reduce the dimensional of a vector of parameters, with the simultaneous goal of maximizing the identification rates.

This procedure has been executed using 5 sentences for each speaker. Figure 5 shows the obtained results with the database described previously, using the same sentences to obtain the model sizes and the identification rates (Obviously, this is not a real situation, because the model size is set on a posteriori basis). In order to compare the constant model size for all the speakers with the proposed variable model size (that is, a different model size for each speaker), the mean number of quantization bits is computed with the following formulation, assuming that all the speakers are equally probable:

$$\overline{N}_q = \langle N_{qi} \rangle = \log_2 \left( \frac{1}{N} \sum_{i=1}^{N} 2^{N_{qi}} \right)$$

Experimentally we have found the following results:
- The algorithm modifies the initial number of bits of each model size with 3 possibilities: $\{N_{qi},(N_{qi}+1),(N_{qi}+2)\}$, being the less probable the last one. Thus, the algorithm has been executed with several initialization sizes. Figure 5 shows the $\overline{N}_q$ and the identification rate values obtained at each simulation.
- Obviously the distance values obtained for a given codebook size are smaller if the model size is increased. This fact can be check in figure 6. This observation is more important when the number of bits is low than when it is high. In order to compensate this fact, we have normalized the distortions by a factor $\frac{1}{N_q}$.
- The proposed method achieves similar results to the classical fix model sizes with an equivalent mean model size two bits smaller. This is valid for the normal codebook sizes of the classical fix size approach (5 to 7 bits), when de codebook sizes are computed with a posteriori information. Unfortunately, this is not valid if different sentences are used for codebook size calculation and identification rates. For this reason, we will try to propose another algorithm that does not require a posteriori information.

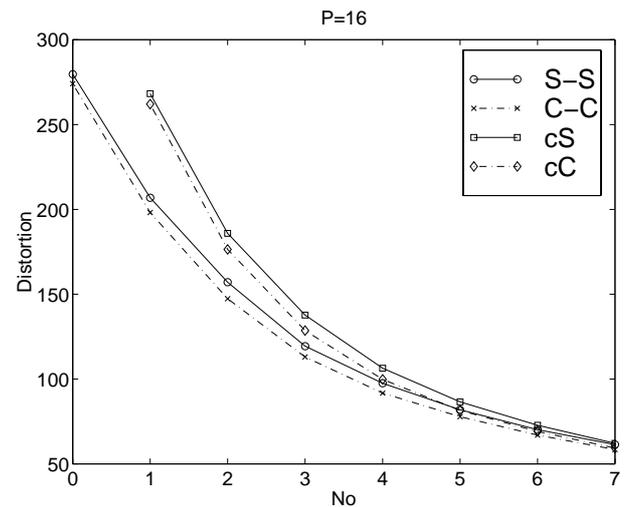

*Figure 6: Quantization distortions as function of Nq*

## 2.3. Proposed algorithm number 2.

In this section we propose a codebook size selection method that does not require a trial and error procedure. The goal is to find a relation between the performance of a given model size and the requiered number of quantization bits.

It is based on the quantization distortion and variance observations, and the goal is to use a model that yields similar quantization distortions and variances to the models of the others speakers, with their respective sentences.

Figures 7 and 8 show the histogram of the distances from one test sentence to his model, for all the speakers of the database. It can be seen that for all the range of model size values, several speakers exist whose distortion is more than two times greater, with respect to the mean distortion value.

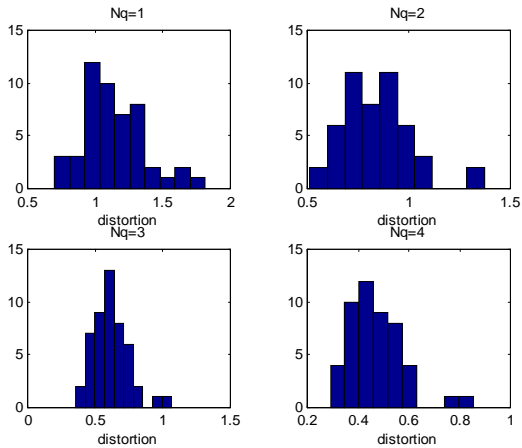

*Figure 7* Histograms of the quantization distortion (Nq=1 to 4)

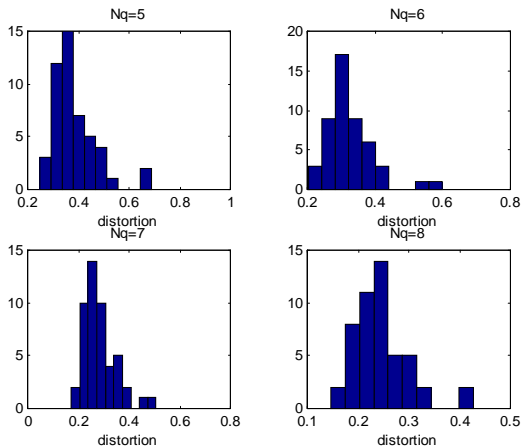

*Figure 8* Histograms of the quantization distortion (Nq=5 to 8)

We have checked that an algorithm based on the quantization performance does not produce good enough results. Taking into account that the discrimination performance is also based on the variance of the distortions, we can define a combined criterion on distortion and variance.

Figure 9 shows the ratio between the standard deviation and the mean of the distortion. Obviously the distortions over test sentences of the same speaker are smaller when the number of quantization bits is increased. This figure has been obtained in two different situations: same language for training and testing, and different languages (mismatch).

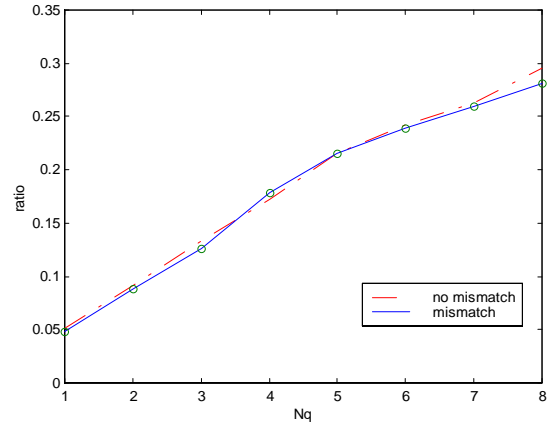

*Figure 9* Ratio between the standard deviation (other speakers' sentences) and the mean of the distortion measures (same speaker sentences).

Based on the observations of figure number 9, a criterion has been defined. We propose the following algorithm, that tries to find a straight forward relation between the distortion and standard deviation, and the required number of quantization bits:

- Those speakers whose ratio between standard deviation and mean distortion is smaller that the mean of this ratio obtained averaging the value for all the speakers, increase the number of bits of his model.

This second algorithm is not an iterative procedure. Thus, it is faster to obtain the bit assignment than algorithm number 1.

Figure 10 shows the identification rates using this algorithm. This figure has been obtained increasing in one bit the model size of those speakers whose ratio between standard deviation and the distortion mean is smaller than 50% of the mean ratio for all the speakers.

It seems that this criterion is not good enough, or at least it can not obtain similar results to the proposed algorithm number 1.

Thus, although intuitively it seems that more parameters are needed for speakers with a more complicated occupancy of the space parameter (requiring a model with more parameters than the "normal speakers", in order to obtain an accurate model), it is not a trivial question which criterion must be used.

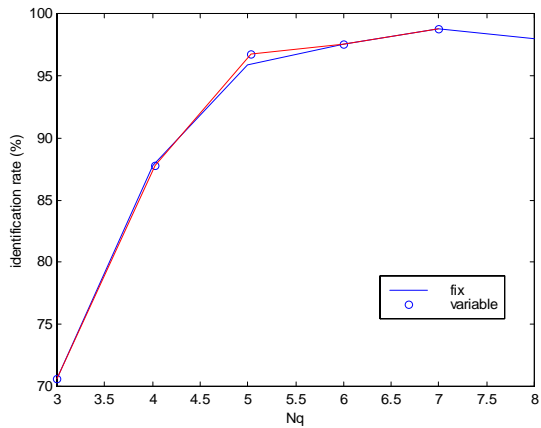

*Figure 10 Identification rates with classical VQ and algorithm 2.*

## 3. CONCLUSIONS

The most important conclusions are the following ones:
- An iterative algorithm (section 2.2) can improve the recognition rates, increasing the model size of several speakers. Unfortunately this is only true when a posteriori algorithm is used (the codebook size is selected using the same test sentences used for the final identification rates).
- The ratio between standard deviation and mean distortion is not the unique important parameter for the model size selection.
- We believe that the use of different model size for each speaker can improve the identification rates, and that model size is as important as other application fields, such as speech recognition (different models or model sizes are used for each phoneme), polynomial interpolation, neural networks, etc.
- Although our algorithm does not improve the results without using a posteriori information, we believe that significative results can be obtained with a steepest study and more statistical information (a higher number of sentences for setting up the codebook sizes).

**Acknowledgements**
I want to acknowledge the useful conversations about pattern recognition and the speaker's model size, with Dr. Enric Monte-Moreno, from the Politecnical University of Catalunya.